\documentclass{cjpsuplf}    
\usepackage{graphicx}
\begin{document}
\title{Beyond the standard model physics at RHIC \\in polarized pp collision
}
\authori{Jiro Murata}
\addressi{RIKEN, Wako 351-0198, Japan}
\authorii{}     
\addressii{}
\authoriii{}     
\addressiii{}
\headtitle{Beyond the standard model physics at RHIC in polarized pp collision\ldots}
\headauthor{Jiro Murata}  
\specialhead{Jiro Murata: Beyond the standard model physics at RHIC in
polarized pp collision \ldots}
\evidence{}
\daterec{}    
\suppl{A}  \year{2002}
\setcounter{page}{1}
\maketitle

\begin{abstract}
A polarized hadron collider experiment must have a great discovery potential
for a search of physics beyond the standard model.
Experimental data of various symmetry tests at RHIC are going to be
obtained within a few years.
The author developed a simulation tool, studying a sensitivity of
hunting contact interaction at RHIC by measuring parity violating 
spin asymmetries.
\end{abstract}

\section{Introduction}     
One of the most vital tasks of an experiment using a polarized hadron
collider in a new energy region is the study of physics beyond the standard model.
In the past few years, several theoretical works
\cite{Mike,Soffer,Taxil} have been devoted to the studies of the contact
interaction (or, compositeness) \cite{Eichten} and new gauge bosons 
$W'^{\pm}$ \cite{Virey-W} and $Z'$ \cite{Virey-Z}. In these theoretical works, \cite{Virey-Jet} it has been
shown that RHIC can reach a similar sensitivity to that of the TEVATRON, due to its polarized beam.

\begin{figure}[h]
\begin{center}
\includegraphics[width=0.3\linewidth]{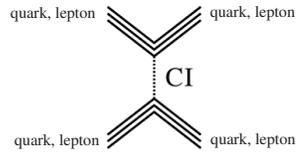}
\end{center}
\caption{Possible Feynmann Diagram of the contact
 interaction with energy scale $\Lambda$. }
\label{cifmdy}
\end{figure}

The purpose of the present study is to explore the discovery potential
for physics beyond the standard model from the experimental perspective. 
Considering the family pattern of quarks and leptons, 
compositeness is an fascinating idea.
For there are no standard sub-quark models, model independent treatment
using a compositeness scale $\Lambda$ is often used.
We call such new generalized interaction as the contact interaction.
One possible Feynmann diagram is shown in Fig.\ref{cifmdy}.
For any kind of an new interaction with energy scale $\Lambda$ can 
be taken into account, for example, new gauge boson exchange between
``standard'' quarks can be understood as a kind of the contact interaction.
$\Lambda$ is a model-independent scale parameter of the contact
interaction, defined as $F(Q^2)=(1+Q^2/\Lambda^2)^{-1}$, 
where $F(Q^2)$ is a `form factor' of the quark and lepton.
Therefore, sometimes one interpret the physical meanings of the the contact
interaction as quark (lepton) compositeness.
However, the contact interaction has general formalism, so any kind
of new interaction can be roughly included.

\section{Simulation Procedure using P{\small YTHIA}+P{\small OLBY}}
In order to study  physics sensitivity and to determine a search window,
simulation studies with an event generator are indispensable. 
Several non-standard scenarios can be examined using P{\small YTHIA}. 
The contact interaction can be also examined through it. 
However, as with all other subprocesses, P{\small YTHIA} includes only 
helicity-averaged cross sections for the contact interaction.
The author developed a plug-in program `P{\small OLBY}'
for P{\small YTHIA} by which hadronic spin asymmetries for Drell-Yan and 
quark-scattering process can be examined. 
Helicity-selected matrix elements have been theoretically obtained in LO
not only for the standard model but also for the contact interaction.
In case of Drell-Yan process, corresponding matrix elements were
obtained from Virey's calculation for DIS by crossing \cite{Virey-HERA}.
A weighted mean method is applied in order to extract spin asymmetries
using ``unpolarized'' event generator P{\small YTHIA}.
A weight factor is calculated for each generated event, then its mean
value of the weight factor distribution gives us the final spin
asymmetries.
First, partonic cross section $\hat{\sigma}$ can be obtained at a
kinematics region by summing up the matrix elements for all the possible 
interactions.
In order to estimate the hadronic spin asymmetries, we need hadronic cross
sections $\sigma^{h_1 h_2}$ with helicity $h_1, h_2$ selection.
They can be obtained by combining with polarized quark distribution
functions $q^{\pm}$. For example, $\sigma^{++}$ can be obtained as; 
\begin{equation}
\sigma^{++}=
\hat{\sigma}^{++}q_1^+ q_2^+
+\hat{\sigma}^{--}q_1^- q_2^-
+\hat{\sigma}^{+-}q_1^+ q_2^-
+\hat{\sigma}^{-+}q_1^- q_2^+ .
\end{equation}
Using the hadronic cross sections, we can estimate the weight factor in 
combinations between different spin selected hadronic cross sections.
For single spin asymmetry 
\begin{equation}
A_L=\frac{\sigma^{-}-\sigma^{+}}{\sigma^{-}+\sigma^{+}}
=\frac{\sigma^{--}+\sigma^{-+}-\sigma^{++}-\sigma^{+-}}{\sigma^{--}+\sigma^{-+}+\sigma^{++}+\sigma^{+-}}
\end{equation}
and parity violating double spin asymmetry
\begin{equation}
A^{PV}_{LL}=\frac{\sigma^{--}-\sigma^{++}}{\sigma^{--}+\sigma^{++}} ,
\end{equation}
corresponding weight factors can be obtained by making the same cross
section ratio in event-by-event.
The distribution of the weight factor depends on the quark distribution functions, therefore,
the process of making a mean value on these distribution implies an
Monte-Carlo integration over the quark distribution functions.
There are two free parameters in the matrix elements for the contact
interaction. First one is the scale parameter $\Lambda$ and the second
one is a sign of the interference between the standard model and the
contact interaction.
\newpage
\section{Drell-Yan process}
\begin{figure}[h]
\begin{center}
\includegraphics[width=0.9\linewidth]{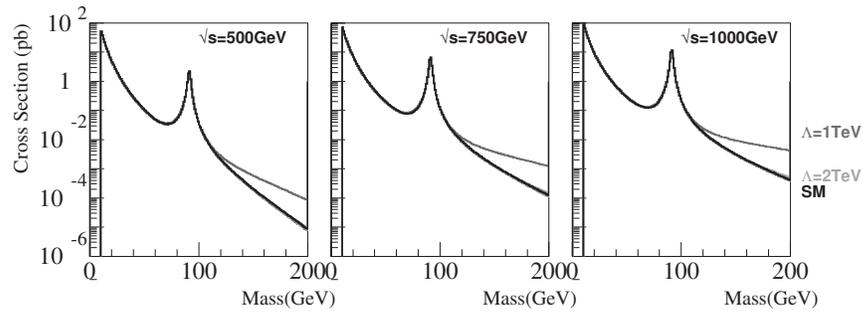}
\end{center}
\caption{Cross section of Drell-Yan process is shown including the contact
 interaction.}
\label{dycs}
\end{figure}

Parity violating spin asymmetries are examined using P{\small
YTHIA}+P{\small OLBY}. 
In P{\small YTHIA}, Drell-Yan events can be
generated in single subprocess, which include the standard model
$\gamma^*/Z^0$ exchange and also the contact interaction with its scale $\Lambda$.
Fig. \ref{dycs} shows the results on Drell-Yan yield with
selections of beam energies ($\sqrt{s}$ = 500, 750, 1000 GeV) and the scale
$\Lambda$. Here, GS-A polarized parton distribution function \cite{GS} is used.
At any beam energy, excess from the standard model are observed
when we require $\Lambda$ = 1 TeV, however, the excess is negligible if 
$\Lambda\geq$ 2 TeV. Considering the percent-level systematic error
in a cross section measurement, observation of the cross section excess 
is hopeless. 
In Drell-Yan process, contribution from the pure contact interaction is very 
small because of its $\Lambda$ dependence of $\Lambda^{-4}$, 
on the other hand, the interference terms between
the contact interaction and $Z^0$, $\gamma^*$ have $\Lambda^{-2}$ dependence.

\begin{figure}[h]
\begin{center}
\includegraphics[width=0.6\linewidth]{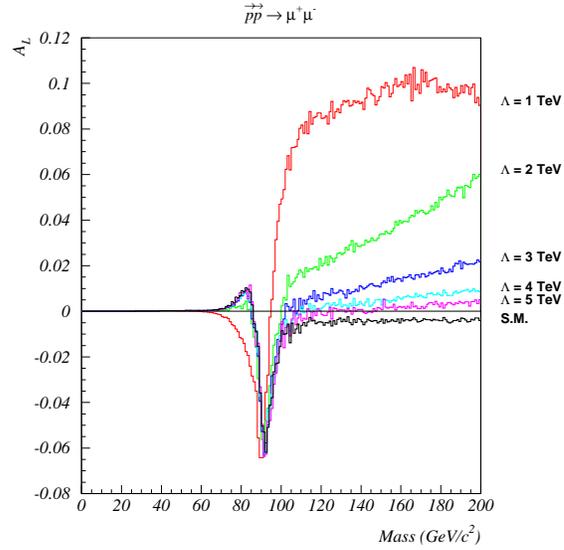}
\end{center}
\caption{Single spin asymmetries $A_L$ are drawn as functions of
 Drell-Yan mass with selections of $\Lambda$.}
\label{asymal}
\end{figure}

\begin{figure}[h]
\begin{center}
\includegraphics[width=0.6\linewidth]{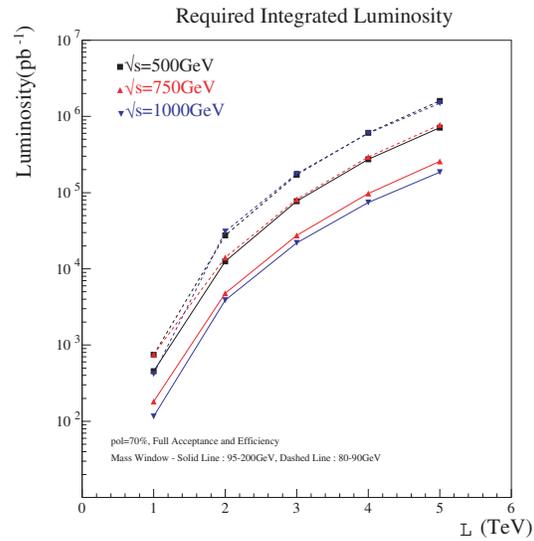}
\end{center}
\caption{Required integrated luminosity as functions of $\Lambda$ is
 shown.}
\label{lumidy}
\end{figure}

On the other hand, a polarized experiment has a great
advantage because of its systematic error cancellation. Simulation
results using P{\small YTHIA}+P{\small OLBY} on the parity 
violating single spin asymmetry $A_L$ are shown in Fig. \ref{asymal}.
Pure $Z^0$ exchange and $\gamma^*-Z^0$ interference contribute on the
parity violation on the standard model. We can see the large deviation
from the standard model if we include the contact interaction. Dominant
contribution on the parity violation is the interference term between
the contact interaction and the standard model. 
In Fig.\ref{asymal}, we can see a clear deviation from the standard
model even at $\Lambda$ = 5 TeV, however, the experimental statistical error
have to be also large. Because of its relatively small cross section, required 
integrated luminosity is large as shown in Fig.\ref{lumidy}. 
Considering the RHIC luminosity of 800 pb$^{-1}$ at $\sqrt{s}=500$GeV, 
our sensitivity is limited around $\Lambda \sim 1$ TeV, which is lower than
the already established limits in $e^+ e^-$ collisions for the
electron-quark contact interaction ($\Lambda>5.4$ TeV) by ALEPH and
in $\bar p p$ collisions for the muon-quark contact interaction 
($\Lambda>2.9$ TeV) by CDF \cite{PDG}.

\section{Inclusive Jet production process}
\begin{figure}[h]
\begin{center}
\includegraphics[width=0.3\linewidth]{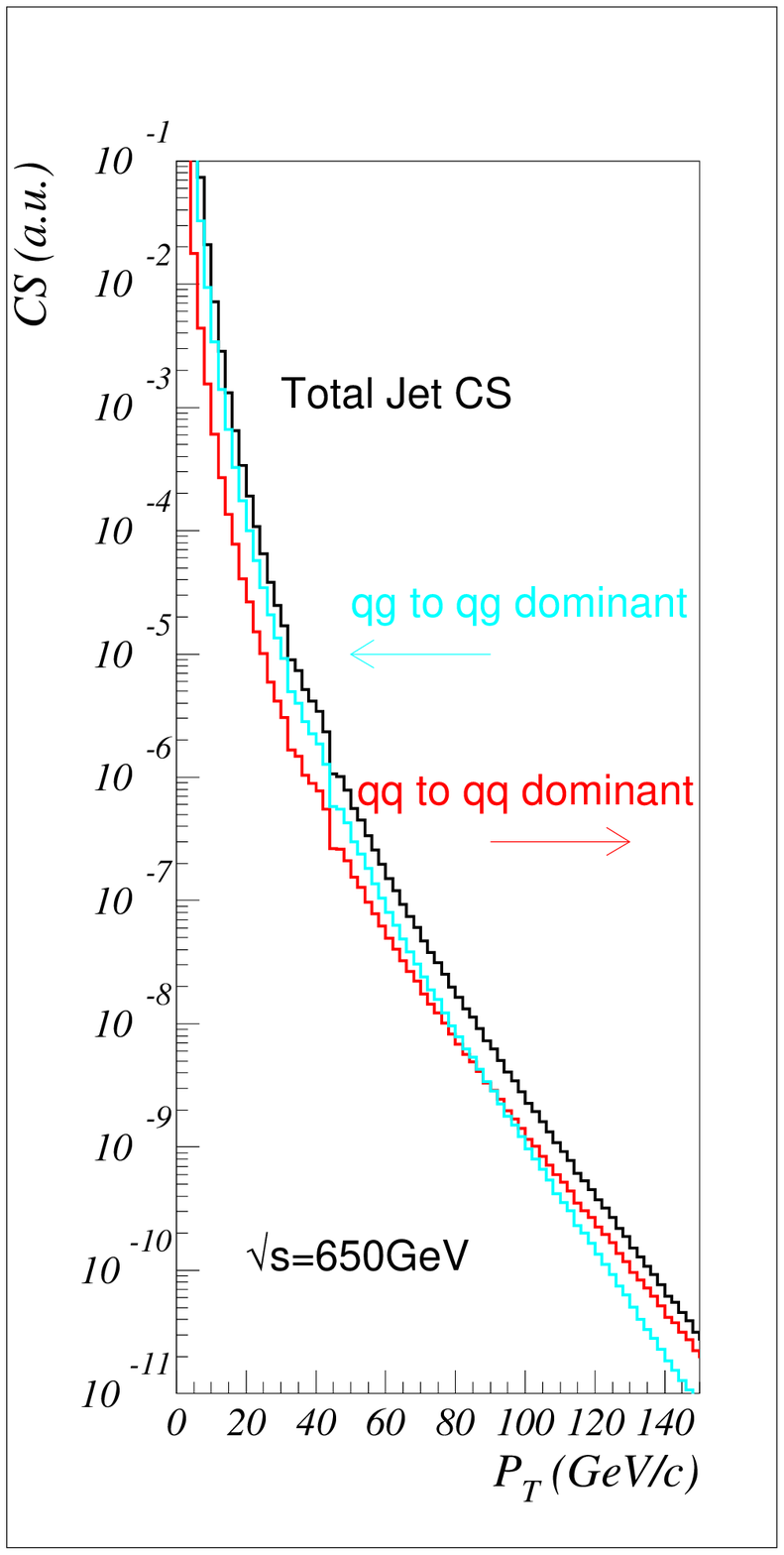}
\end{center}
\caption{Jet production cross section is shown. At high Pt region,
 quark-quark scattering is dominant. Small bump around $M_W/2$ corresponds
Jacobean Peak}
\label{jetcs}
\end{figure}

\begin{figure}[h]
\begin{center}
\includegraphics[width=0.45\linewidth]{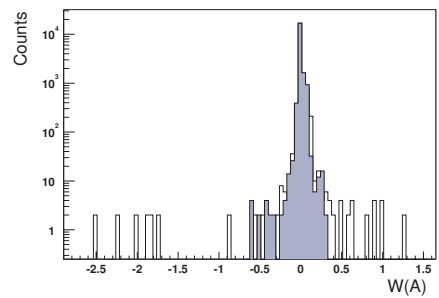}
\end{center}
\caption{Sample plot of the weight factor. Hatched(Blank) histogram
 shows before(after) the interference correction.}
\label{weight}
\end{figure}

\begin{figure}[h]
\begin{center}
\includegraphics[width=\linewidth]{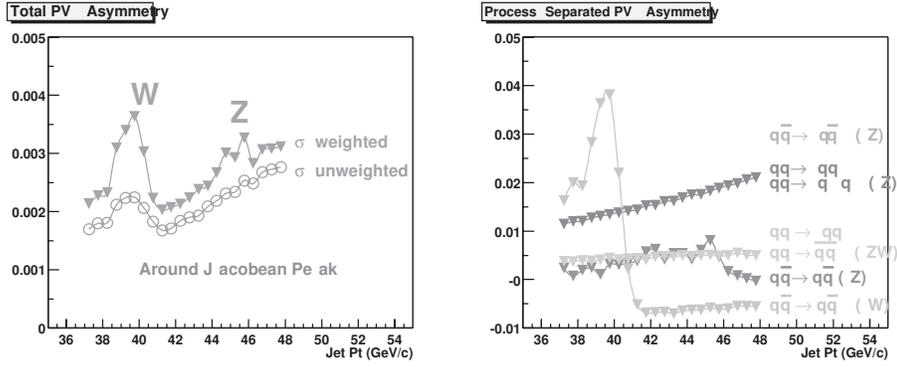}
\end{center}
\caption{Fine structure of the Jacobean peak on parity violating spin asymmetries.}
\label{jetjacob}
\end{figure}

\begin{figure}[h]
\begin{center}
\includegraphics[width=0.5\linewidth]{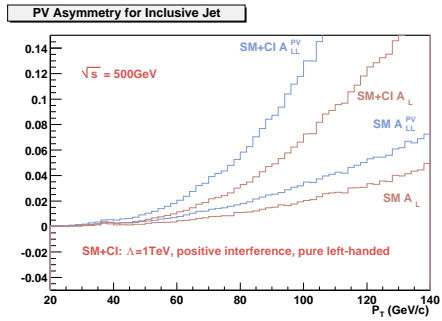}
\end{center}
\caption{Expected parity violating spin asymmetries.}
\label{pvjetc}
\end{figure}

\begin{figure}[h]
\begin{center}
\includegraphics[width=0.5\linewidth]{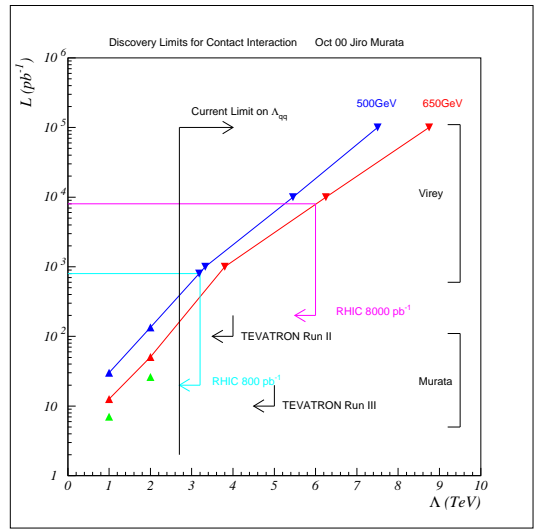}
\end{center}
\caption{Required integrated luminosities are plotted as functions of $\Lambda$.}
\label{lumijet}
\end{figure}

Although Drell-Yan process is simple and clean, we need to examine other
process, which has large production cross section because of the small
yield on Drell-Yan process.
Parity violation in the inclusive jet production process has already examined by J.-M. Virey 
theoretically \cite{Virey-Jet}. 
The jet production cross section is shown in Fig.\ref{jetcs} as
functions of jet $P_T$. 
Lower $P_T$ region is dominated by gluon-related processes, on the other
hand, higher $P_T$ region is dominated by quark-quark scattering. 
The contact interaction couples to leptons and quarks, but not to gluons in
the current model. 
It means that we do not think gluon is a composite particle. 
Therefore, sensitive $P_T$ region of the contact interaction must be
limited to the higher $P_T$ region.  
In case of Drell-Yan process, all the interactions we need to consider, are
the standard model $\gamma^*/Z^0$, the contact interaction, and their
interferences.
On the other hand, we need to take into consideration
QCD, $\gamma^*$, $Z^0$, $W^\pm$, the contact interaction, and their complex
interference terms for the jet production.
In P{\small YTHIA}, above processes must be generated as independent 
subprocesses. 
This restriction cause severe problem when we consider the interference
terms between different subprocesses in P{\small YTHIA}.
It is because there cannot exist interference effects if we distinguish
the subprocess.
In case of Drell-Yan process, we can use single subprocess, therefore, the
interference effects can be automatically included.
In order to restore quantum mechanics in the jet production, a special
treatment is required.
In P{\small OLBY}, a hadronic spin asymmetry $A$ is estimated using
event-by-event asymmetry weight factor $W(A)$;
\begin{equation}
A=\frac{\sum_{i=event}W^i(A)}{\sum_{i=event}} .
\label{wa}
\end{equation}
Here, the ``events'' are generated by P{\small YTHIA} subprocess.
The ``interference correction'' can be done by
\begin{equation}
W^i(A)\rightarrow W^i(A)\times 
\frac{\sum_{process=QCD,Weak,EM,CT} \sigma ^{process}}{\sigma^{QCD}}
\label{intcor}
\end{equation}
with Eq.\ref{wa}.
Then if we use pure QCD subprocess for the P{\small YTHIA} event
generation, contribution from other processes on the asymmetry can be
taken into account by summing up over all the processes.
Fig.\ref{weight} shows a sample distribution of $W(A)$ before/after the
interference correction of Eq.\ref{intcor}. The mean value
of the $W(A)$ distribution shows the final hadronic spin asymmetry.
By the operation Eq.\ref{intcor}, $W(A)$ distribution is distorted
and we can get the corrected asymmetry from the distorted distribution.
By using Eq.\ref{intcor}, we can estimate accurate asymmetries.
For an example, parity violation around Jacobean peak is examined.
In Fig.\ref{jetjacob}, we can confirm the large contribution from 
the interference term on the parity violation around the Jacobean peak. 
Subprocess-separated plot in Fig.\ref{jetjacob} shows that 
$q\bar{q}'\rightarrow q\bar{q}'$
subprocess dominates the parity violation at the W peak.
Fig.\ref{pvjetc} shows a preliminary results, which can be compared with
the theoretical calculation \cite{Virey-Jet}. Considering the difference on
the pseudo rapidity cut and binding, the results meets well with the
calculation by Virey.
Using these results, discovery potential at RHIC can be explored. In
Fig.\ref{lumijet}, required integrated luminosities are plotted as
functions of $\Lambda$.
Here combined results from the present study at lower $\Lambda$ region
and that from Virey at higher $\Lambda$ region \cite{Virey-650} are shown.
It is shown that we can reach more than 3 TeV sensitivity, which is
higher than the current limit of 2.7 TeV reported by D0 for the
quark-quark contact interaction, \cite{PDG} with the current RHIC plan. 
It is also shown that the sensitivity limit can be more than 6 TeV 
using the RHIC upgrade plan. 
We can calculate many combinations of parity violating spin asymmetries
using P{\small YTHIA}+P{\small OLBY} with many realistic experimental 
conditions. For the procedure using weighted mean can be understood as a
kind of Monte-Carlo integration, the calculation speed is very fast.
Therefore this is a strong tool to examine an experimental sensitivity and
also to make a standard model reference after getting experimental data.

However, in order to use it as a standard model reference, NLO calculation
is indispensable. NLO effect is difficult to take into an event
generator. The situation is similar to the treatment of the
interference. One large assumption for the interference correction is
that, pure QCD process (generated events) and other processes must have their
same final states. In principle, NLO correction using similar cross
section correction like;
\begin{equation}
W^i(A)\rightarrow W^i(A)\times 
\frac{\sum_{process=NLO+LO} \sigma ^{process}}{\sigma^{LO}}
\label{nlocor}
\end{equation}
should be able to be applied.
Theoretical calculations in NLO are now under progress \cite{NLO-G}. We should and will be
able to be ready to examine experimental data from RHIC in near future.

\section*{Acknowledgement}

{\small Many thanks are due to J.-M. Virey and J. Soffer for the
continuous help from theoretical side, especially for the treatment of
the matrix elements. The author is also grateful to G. Bunce, N. Saito
and M. Tannenbaum for the advises on the experimental aspects and on the simulation formalism. 
}

\section{Future perspectives}

\end{document}